\documentclass[twocolumn]{aastex63}
\usepackage{amsmath}
\usepackage{upgreek}
\usepackage[encapsulated]{CJK} 
\usepackage{ucs}
\newcommand{\cntext}[1]{\begin{CJK}{UTF8}{gbsn}#1\end{CJK}}

\providecommand{\sorthelp}[1]{}

\received{July 16, 2021}
\revised{November 8, 2021}
\accepted{November 12, 2021}
\published{February 9, 2022}
\submitjournal{ApJ}

\shorttitle{The CLASS multi-frequency on-sky receiver performance}
\shortauthors{Dahal et al.}

\graphicspath{{./}{figures/}}

\begin{document}

\title{Four-year Cosmology Large Angular Scale Surveyor (CLASS) Observations: On-sky Receiver Performance at 40, 90, 150, and 220 GHz Frequency Bands}

\correspondingauthor{Sumit Dahal}
\email{sumit.dahal@nasa.gov}

\author[0000-0002-1708-5464]{Sumit Dahal}
\affiliation{NASA Goddard Space Flight Center, 8800 Greenbelt Road, Greenbelt, MD 20771, USA}
\affiliation{Department of Physics and Astronomy, Johns Hopkins University, 3701 San Martin Drive, Baltimore, MD 21218, USA}

\author[0000-0002-8412-630X]{John~W. Appel}
\affiliation{Department of Physics and Astronomy, Johns Hopkins University, 3701 San Martin Drive, Baltimore, MD 21218, USA}

\author[0000-0003-3853-8757]{Rahul Datta}
\affiliation{Department of Physics and Astronomy, Johns Hopkins University, 3701 San Martin Drive, Baltimore, MD 21218, USA}

\author{Michael K. Brewer}
\affiliation{Department of Physics and Astronomy, Johns Hopkins University, 3701 San Martin Drive, Baltimore, MD 21218, USA}

\author[0000-0001-7941-9602]{Aamir Ali}
\affiliation{Department of Physics, University of California, Berkeley, CA 94720, USA}
\affiliation{Department of Physics and Astronomy, Johns Hopkins University, 3701 San Martin Drive, Baltimore, MD 21218, USA}

\author[0000-0001-8839-7206]{Charles L. Bennett}
\affiliation{Department of Physics and Astronomy, Johns Hopkins University, 3701 San Martin Drive, Baltimore, MD 21218, USA}

\author[0000-0001-8468-9391]{Ricardo Bustos}
\affiliation{Facultad de Ingenier\'{i}a, Universidad Cat\'{o}lica de la Sant\'{i}sima Concepci\'{o}n, Alonso de Ribera 2850, Concepci\'{o}n, Chile}

\author{Manwei Chan}
\affiliation{Department of Physics and Astronomy, Johns Hopkins University, 3701 San Martin Drive, Baltimore, MD 21218, USA}

\author[0000-0003-0016-0533]{David T. Chuss}
\affiliation{Department of Physics, Villanova University, 800 Lancaster Avenue, Villanova, PA 19085, USA}

\author{Joseph Cleary}
\affiliation{Department of Physics and Astronomy, Johns Hopkins University, 3701 San Martin Drive, Baltimore, MD 21218, USA}

\author[0000-0002-0552-3754]{Jullianna D. Couto}
\affiliation{Department of Physics and Astronomy, Johns Hopkins University, 3701 San Martin Drive, Baltimore, MD 21218, USA}

\author{Kevin L. Denis}
\affiliation{NASA Goddard Space Flight Center, 8800 Greenbelt Road, Greenbelt, MD 20771, USA}

\author{Rolando D\"{u}nner}
\affiliation{Instituto de Astrof\'isica and Centro de Astro-Ingenier\'ia, Facultad de F\'isica, Pontificia Universidad Cat\'olica de Chile, Av. Vicu\~na Mackenna 4860, 7820436 Macul, Santiago, Chile}

\author[0000-0001-6976-180X]{Joseph Eimer}
\affiliation{Department of Physics and Astronomy, Johns Hopkins University, 3701 San Martin Drive, Baltimore, MD 21218, USA}

\author{Francisco Espinoza}
\affiliation{Facultad de Ingenier\'{i}a, Universidad Cat\'{o}lica de la Sant\'{i}sima Concepci\'{o}n, Alonso de Ribera 2850, Concepci\'{o}n, Chile}

\author[0000-0002-4782-3851]{Thomas Essinger-Hileman}
\affiliation{NASA Goddard Space Flight Center, 8800 Greenbelt Road, Greenbelt, MD 20771, USA}
\affiliation{Department of Physics and Astronomy, Johns Hopkins University, 3701 San Martin Drive, Baltimore, MD 21218, USA}

\author{Joseph E. Golec}
\affiliation{Department of Physics, University of Chicago, Chicago, IL 60637, USA}

\author[0000-0003-1248-9563]{Kathleen Harrington}
\affiliation{Department of Astronomy and Astrophysics, University of Chicago, 5640 South Ellis Avenue, Chicago, IL 60637, USA}
\affiliation{Department of Physics and Astronomy, Johns Hopkins University, 3701 San Martin Drive, Baltimore, MD 21218, USA}

\author[0000-0001-9238-4918]{Kyle Helson}
\affiliation{NASA Goddard Space Flight Center, 8800 Greenbelt Road, Greenbelt, MD 20771, USA}
\affiliation{University of Maryland Baltimore County, 1000 Hilltop Circle, Baltimore, MD 21250, USA}

\author[0000-0001-7466-0317]{Jeffrey Iuliano}
\affiliation{Department of Physics and Astronomy, Johns Hopkins University, 3701 San Martin Drive, Baltimore, MD 21218, USA}

\author{John Karakla}
\affiliation{Department of Physics and Astronomy, Johns Hopkins University, 3701 San Martin Drive, Baltimore, MD 21218, USA}

\author[0000-0002-4820-1122]{Yunyang Li (\cntext{李云炀}\!\!)}
\affiliation{Department of Physics and Astronomy, Johns Hopkins University, 3701 San Martin Drive, Baltimore, MD 21218, USA}

\author[0000-0003-4496-6520]{Tobias~A. Marriage}
\affiliation{Department of Physics and Astronomy, Johns Hopkins University, 3701 San Martin Drive, Baltimore, MD 21218, USA}

\author{Jeffrey~J. McMahon}
\affiliation{Department of Astronomy and Astrophysics, University of Chicago, 5640 South Ellis Avenue, Chicago, IL 60637, USA}
\affiliation{Department of Physics, University of Chicago, Chicago, IL 60637, USA}

\author[0000-0002-2245-1027]{Nathan~J. Miller}
\affiliation{Department of Physics and Astronomy, Johns Hopkins University, 3701 San Martin Drive, Baltimore, MD 21218, USA}

\author{Sasha Novack}
\affiliation{Department of Physics and Astronomy, Johns Hopkins University, 3701 San Martin Drive, Baltimore, MD 21218, USA}

\author[0000-0002-5247-2523]{Carolina N\'{u}\~{n}ez}
\affiliation{Department of Physics and Astronomy, Johns Hopkins University, 3701 San Martin Drive, Baltimore, MD 21218, USA}

\author[0000-0003-2838-1880]{Keisuke Osumi}
\affiliation{Department of Physics and Astronomy, Johns Hopkins University, 3701 San Martin Drive, Baltimore, MD 21218, USA}

\author[0000-0002-0024-2662]{Ivan L. Padilla}
\affiliation{Department of Physics and Astronomy, Johns Hopkins University, 3701 San Martin Drive, Baltimore, MD 21218, USA}

\author[0000-0002-5890-9554]{Gonzalo~A. Palma}
\affiliation{Departamento de F\'isica, FCFM, Universidad de Chile, Blanco Encalada 2008, Santiago, Chile}

\author[0000-0002-8224-859X]{Lucas Parker}
\affiliation{Space and Remote Sensing, MS D436, Los Alamos National Laboratory, Los Alamos, NM 87544, USA}
\affiliation{Department of Physics and Astronomy, Johns Hopkins University, 3701 San Martin Drive, Baltimore, MD 21218, USA}

\author[0000-0002-4436-4215]{Matthew~A. Petroff}
\affiliation{Department of Physics and Astronomy, Johns Hopkins University, 3701 San Martin Drive, Baltimore, MD 21218, USA}

\author[0000-0001-5704-271X]{Rodrigo Reeves}
\affiliation{CePIA, Departamento de Astronom\'{i}a, Universidad de Concepci\'{o}n, Concepci\'{o}n, Chile}

\author{Gary Rhoades}
\affiliation{Department of Physics and Astronomy, Johns Hopkins University, 3701 San Martin Drive, Baltimore, MD 21218, USA}

\author[0000-0003-4189-0700]{Karwan Rostem}
\affiliation{NASA Goddard Space Flight Center, 8800 Greenbelt Road, Greenbelt, MD 20771, USA}

\author[0000-0003-3487-2811]{Deniz A. N. Valle}
\affiliation{Department of Physics and Astronomy, Johns Hopkins University, 3701 San Martin Drive, Baltimore, MD 21218, USA}

\author[0000-0002-5437-6121]{Duncan J. Watts}
\affiliation{Institute of Theoretical Astrophysics, University of Oslo, P.O. Box 1029 Blindern, N-0315 Oslo, Norway}
\affiliation{Department of Physics and Astronomy, Johns Hopkins University, 3701 San Martin Drive, Baltimore, MD 21218, USA}

\author[0000-0003-3017-3474]{Janet L. Weiland}
\affiliation{Department of Physics and Astronomy, Johns Hopkins University, 3701 San Martin Drive, Baltimore, MD 21218, USA}

\author[0000-0002-7567-4451]{Edward J. Wollack}
\affiliation{NASA Goddard Space Flight Center, 8800 Greenbelt Road, Greenbelt, MD 20771, USA}

\author[0000-0001-5112-2567]{Zhilei Xu (\cntext{徐智磊}$\!\!$)}
\affiliation{ MIT Kavli Institute, Massachusetts Institute of Technology, 77 Massachusetts Avenue, Cambridge, MA 02139, USA}
\affiliation{Department of Physics and Astronomy, Johns Hopkins University, 3701 San Martin Drive, Baltimore, MD 21218, USA}

\begin{abstract}
The Cosmology Large Angular Scale Surveyor (CLASS) observes the polarized cosmic microwave background (CMB) over the angular scales of 1$^\circ \lesssim \theta \leq$ 90$^\circ$ with the aim of characterizing primordial gravitational waves and cosmic reionization. We report on the on-sky performance of the CLASS Q-band (40~GHz), W-band (90 GHz), and dichroic G-band (150/220 GHz) receivers that have been operational at the CLASS site in the Atacama desert since June 2016, May 2018, and September 2019, respectively. We show that the noise-equivalent power measured by the detectors matches the expected noise model based on on-sky optical loading and lab-measured detector parameters. Using Moon, Venus, and Jupiter observations, we obtain power-to-antenna-temperature calibrations and optical efficiencies for the telescopes. From the CMB survey data, we compute instantaneous array noise-equivalent-temperature sensitivities of 22, 19, 23, and 71 $\mathrm{\upmu}$K$_\mathrm{cmb}\sqrt{\mathrm{s}}$ for the 40, 90, 150, and 220~GHz frequency bands, respectively. These noise temperatures refer to white noise amplitudes, which contribute to sky maps at all angular scales. Future papers will assess additional noise sources impacting larger angular scales.
\end{abstract}

\keywords{\href{http://astrothesaurus.org/uat/322}{Cosmic microwave background radiation (322)}; \href{http://astrothesaurus.org/uat/435}{Early Universe (435)}; \href{http://astrothesaurus.org/uat/1146}{Observational Cosmology (1146)}; \href{http://astrothesaurus.org/uat/799}{Astronomical instrumentation (799)}; \href{http://astrothesaurus.org/uat/1277}{Polarimeters (1277)}; \href{http://astrothesaurus.org/uat/259}{CMBR Detectors (259)}}

\section{Introduction} \label{sec:intro}
The cosmic microwave background (CMB) polarization is a unique probe to study the origin and evolution of the universe. The E-mode polarization component with $(-1)^\ell$ parity constrains the optical depth to reionization $\tau$, which is currently the least constrained fundamental $\Lambda$CDM parameter \citep{bennett2013,planck2016-l06,watts2018}. This uncertainty in $\tau$ will be a limiting factor for improving constraints on the sum of neutrino masses from future CMB experiments \citep{cmbs42016}. Additionally, the B-mode polarization component with $(-1)^{\ell+1}$ parity  provides constraints on primordial gravitational waves \citep{kamionkowski1997, seljak1997}, which would provide evidence for inflation \citep{starobinski1979, guth1981, linde1982}. However, detecting primordial B-modes (if they in fact exist) is difficult in part because the B-mode signal is much fainter than the polarized Galactic foregrounds. The synchrotron emission from relativistic electrons accelerated in the Galactic magnetic field dominates at frequencies below $\sim$ 70~GHz, while thermal emission from dust grains in the interstellar medium dominates at higher frequencies \citep{bennett2013,watts2015,planck2016-l04}. Therefore, it is necessary to pursue multi-frequency observations for polarized foreground flux characterization and removal.

On small angular scales ($\theta \lesssim 1^{\circ}$), the conversion of brighter E-modes into B-modes through weak gravitational lensing by matter along the line of sight is an additional contaminant that needs to be well characterized, and has been measured by multiple surveys \citep{bicep2016,act2017,polarbear2017,sayre2020}. At large angular scales, the putative primordial B-mode signal has characteristic recombination and reionization peaks at $30 \lesssim \ell \lesssim 200 $ and $\ell \lesssim 10$, respectively. Hence, measurements of the B-mode spectrum on larger than a degree angular scales ($\ell \lesssim 180$) not only help to minimize the contamination from lensing but also are essential to confirm the primordial origin of any potential signal measured at small angular scales. Currently, the tightest constraint on the amplitude of the gravitational waves parameterized through the tensor-to-scalar ratio  $r < 0.036$ comes from combining the \textit{Planck} and WMAP data with the BICEP/\textit{Keck} measurement of the recombination peak \citep{BICEP2021}. Current experiments targeting the reionization peak include PIPER \citep{piper2016}, QUIJOTE \citep{quijote2014}, LSPE \citep{lspe2012}, GroundBIRD \citep{groundbird2012}, and the Cosmology Large Angular Scale Surveyor (CLASS; \citealt{katie2016}).

CLASS maps the CMB polarization at multiple frequencies and targets both the recombination and reionization peaks from the Atacama desert with the aim of measuring a primordial B-mode signal at a sensitivity level of $r \sim$~0.01 and making a near cosmic-variance-limited measurement of $\tau$ \citep{tom2014,katie2016,watts2015,watts2018}. The CLASS 40~GHz (Q band) telescope \citep{eimer2012, appel14}, which helps to measure foreground synchrotron emission, has been operational since June 2016. The dichroic 150/220~GHz (G band) receiver \citep{dahal2020}, which aids in characterizing the polarized Galactic dust emission, started observations in September 2019. CLASS is designed to be most sensitive at 90~GHz (W band) with two telescopes \citep{dahal2018, iuliano2018} optimized for CMB observations near the minimum of polarized Galactic emission. The first W-band telescope  has been operational since May 2018, while the second is planned to be deployed in 2022. All four CLASS telescopes share a similar design \citep{eimer2012, iuliano2018} with a variable-delay polarization modulator (VPM; \citealt{katie2018}) as the first optical element to place the signal band at $\sim$10~Hz, away from the low-frequency 1/f noise.

The CLASS Q-band on-sky performance, optical characterization, circular polarization sensitivity, and instrument stability based on the data taken between June 2016 and March 2018 (``Era 1'') have already been published in a series of papers \citep{appel19,xu2020, petroff2020,padilla2020,harrington2021}. This paper presents the on-sky performance of the Q-band instrument after April 2018 \mbox{(``Era 2'')} upgrades until March 2020, and the W-band and G-band instruments from their respective deployments in May 2018 and September 2019 to March 2020. The Era 2 data presented in this paper correspond to 636~days of observations for the Q-band (411~days with a grille filter installed; see Section \ref{sec:det_Q}), 611~days for the W-band, and 167~days for the G-band receivers. In Section \ref{sec:detectors}, we describe the Q-band instrument upgrades and the W-band and G-band focal plane detector arrays. Section~\ref{sec:opt_band_load} shows the optical passband measurements and the on-sky optical loading extracted from $I$-$V$ measurements. In Section~\ref{sec:noise}, we report the noise performance based on the power spectral density (PSD) of the time-ordered data (TOD). Finally, in Section~\ref{sec:planet}, we report the temperature calibrations of the instruments obtained from dedicated Moon/planet observations and the CMB sensitivities calculated from on-sky data for different CLASS detector arrays.

\begin{deluxetable*}{lcrrrr}[ht]
\tablecaption{ \label{tab:bol_summary} Summary of Median TES Bolometer Parameters}
\tablehead{
\colhead{\textbf{Parameter}} & \colhead{\textbf{Symbol [Unit]}} & \colhead{\textbf{40 GHz}} & \colhead{\textbf{90 GHz}} & \colhead{\textbf{150 GHz}} & \colhead{\textbf{220 GHz}}
}
\startdata
Saturation Power & $P_\mathrm{sat}$ [pW] & 6.3 & 18.4 & 35.0 & 43.8 \\  
Optical Loading\tablenotemark{a} & $P_\gamma$ [pW] & 1.2 & 3.6 & 4.4 & 8.5 \\
Optical Time Constant & $\tau_\gamma$ [ms] & 3.4 & 2.1 & 1.5 & 1.4 \\
Thermal Time Constant & $\tau_\varphi$ [ms] & 17 & 7 & 8 & 6 \\
Heat Capacity & $C$ [$\mathrm{pJ \, K^{-1}}$] & 3 & 4 & 5  & 5  \\
Responsivity & $S$ [$\mathrm{\upmu A \, pW^{-1}}$] & -8.2 & -2.9 & -2.3 & -2.2 \\
Thermal Conductance & $G$ [$\mathrm{pW \, K^{-1}}$] & 177 & 452 & 678 & 835\\
Thermal Conductance Constant & $\kappa$ [$\mathrm{nW \, K^{-4}}$] & 13.4  & 24.9 & 18.9  & 21.8\\
Critical Temperature & $T_\mathrm{c}$ [mK] & 149 & 167 & 205 & 209 \\
Normal Resistance & $R_\mathrm{N}$ [m$\Omega$] & 8.2 & 11.0 & 13.8  & 13.9 \\
Shunt Resistance & $R_\mathrm{sh}$ [$\mathrm{\upmu}\Omega$] & 250  & 250 & 200 & 200 \\
TES Loop Inductance & $L$ [nH] & 500 & 300 & 600 & 600 \\
\enddata
\tablenotetext{a}{The median precipitable water vapor (PWV) during the observing campaign was 1.0 mm for the 40 and 90~GHz arrays and 1.2 mm for the 150 and 220 GHz arrays.}
\end{deluxetable*}

\section{Focal Plane Arrays}\label{sec:detectors}
The focal planes for all CLASS telescopes consist of smooth-walled feedhorns \citep{zeng2010} that couple light to polarization-sensitive transition-edge sensor (TES) bolometers through planar orthomode transducers (OMTs; \citealt{wollack2009, denis2009,rostem2016,chuss2016}). Pulse-tube cooled dilution refrigerators \citep{iuliano2018} keep the bath temperature $\lesssim$~60~mK, which is well below the superconducting transition temperature ($\gtrsim$ 150~mK) for CLASS TESs. These TESs are read out through superconducting quantum interference device (SQUID) amplifiers using time-division multiplexing (TDM; \citealt{reintsema2003, battistelli_2008}).

Table \ref{tab:bol_summary} summarizes the median bolometer parameters for all four CLASS frequency bands. These parameters are based on $I$-$V$ curves acquired by first ramping up the detector bias voltage ($V$) to drive the detectors normal and then stepping down the voltage while recording the current response ($I$) of the detectors. Saturation power ($P_\mathrm{sat}$), thermal conductance ($G$), critical temperature ($T_\mathrm{c}$), and normal resistance ($R_\mathrm{N}$) were measured during dark lab tests by capping all the cold stages of the cryostat with metal plates \citep{appel14, dahal2018, dahal2020}. The optical loading ($P_\gamma$; see Section~\ref{sec:loading}) and responsivity ($S = \mathrm{d}I/\mathrm{d}P_\gamma$) are estimated from on-sky $I$-$V$ data. The optical time constant ($\tau_\gamma$) is obtained by fitting detector TODs for a time constant that minimizes the hysteresis of the VPM signal synchronous with the grid-mirror distance \citep{harrington2021, appel19}. The thermal time constant ($\tau_\varphi$) is obtained by multiplying $\tau_\gamma$ with an electrothermal feedback speed-up factor estimated from $I$-$V$ measurements, and the heat capacity ($C$) is the product of $\tau_\varphi$ and $G$ measured for each detector.

\subsection{Q-band Upgrades}\label{sec:det_Q}
The on-sky performance of the Q-band array from its deployment (June 2016) until March 2018 is described in detail in \citet{appel19}. In April 2018, we recovered eight detectors connected to a readout row that was damaged by an electrostatic discharge during deployment. The optically-sensitive TESs connected to the broken readout channels were shifted to neighboring spare readout channels. After this fix we measure nominal $I$-$V$ responses and successfully bias on transition all 72 optically-sensitive TESs in the array. We found two irregular bolometers in the array: one with good optical efficiency but high noise ($\sim$~10$\times$ higher), and another with low optical efficiency (1\%) but typical noise. We removed these from the analysis presented in this paper. These two detectors are not useful for mapping the sky but can be valuable in understanding and tracking systematic effects of the instrument. 

During the Q-band instrument upgrade, we also replaced eight capacitive metal-mesh filters (MMFs; see \citealt[Sec.~6.3]{tom2014}) located at various cryogenic stages of the receiver with a stack of extruded polystyrene foam filters \citep{choi2013} at the 300~K stage. Transmission measurements of these MMFs fabricated by laser delamination of aluminized 12.6 $\upmu$m polypropylene film yielded  higher reflection coefficients than expected from transmission-line models \citep{whitbourn1985, ulrich1967} and performance of similar filters produced using photolithography \citep{ade2006, tom_thesis}. Two MMFs produced by photolithography of 6 $\upmu$m mylar film coated with 30-50 nm of aluminum remain at the 60 K stage of the receiver. The replacement of the eight MMFs with foam filters did not affect the cryogenic performance of the receiver.

The on-sky optical efficiency after the upgrade was measured at 0.54, which is 12\% higher than the 0.48 obtained from the first era of observations \citep{appel19}. The improved four-year absolute temperature calibration based on Venus and Jupiter observations (see Section~\ref{sec:calibration}) accounts for a 5\% increase in our estimate of the instrument's optical efficiency. The additional 7\% optical efficiency improvement is consistent with $\sim$1\% in-band reflection per removed MMF. The in-band optical loading dropped from 1.6~pW to 1.2~pW even though the optical efficiency increased. 
This suggests the MMFs were coupling power onto the detectors from the 60~K cryostat walls and/or the 300~K optics cage. The increase in optical efficiency combined with lower optical loading improved the per-detector sensitivity by $30\%$ from $258~\mathrm{\upmu} \mbox{K} \sqrt{\mathrm{s}}$~\citep{appel19} to $180~\mathrm{\upmu} \mbox{K} \sqrt{\mathrm{s}}$ (Table~\ref{tab:opt_summary}).

The VPM control system was updated, and the baffle and cage enclosures were replaced to accommodate the Q-band and W-band receivers on a single mount. The new electromagnetic environment resulted in increased susceptibility of the Q-band receiver to radio-frequency (RF) noise, in particular to RF signals synchronous to the VPM controller. The W- and G-band detectors do not show similar sensitivity to the VPM RF signals. To improve the Q-band data quality and stability we installed a thin grille (TG) filter at the front of the vacuum window. The TG filter is a 0.51~mm thick brass plate with 5.25~mm diameter circular holes in a 5.75~mm pitch hexagonal packing. The TG filter greatly reduced detector RF noise, improving data quality at the cost of reducing optical efficiency to 0.43, while keeping the detector optical loading at 1.2~pW, and hence decreasing per detector sensitivity to $217~\mathrm{\upmu} \mbox{K} \sqrt{\mathrm{s}}$.
We are actively exploring TG designs with improved transmission that would return the receiver sensitivity to the benchmarks achieved with no TG filter installed. While the data with no TG filter installed was acquired between May 2018 and January 2019, the remaining data from February 2019 until March 2020 were obtained with the TG filter installed.

\subsection{W-band}\label{sec:det_W}
The W-band instrument started observations at the CLASS site in May 2018. The W-band focal plane contains seven identical modules, each consisting of 37 dual-polarization detectors fabricated on 100-mm silicon wafers \citep{dahal2018, rostem2016}. Each W-band pixel has two TES bolometers with each measuring the power in one of the two orthogonal linear polarization states. The total of 518 bolometers in the W-band focal plane are read out using 28 TDM columns, each multiplexing 22 rows of SQUIDs for a total of 616 readout channels. The remaining SQUID channels that are not connected to one of the 518 optically-sensitive bolometers are either used to characterize readout noise and magnetic field pickup or are connected to a TES bolometer without optical coupling to monitor bath-temperature stability. These non-optical bolometers are not considered for analysis in this paper; therefore, we will use ``bolometers'' to refer to only the optically-sensitive bolometers.

Through in-lab characterization, we had reported in \cite{dahal2018} that 426 out of 518 bolometers were functional (i.e., array yield of 82\%). During the deployment, we lost 19 bolometers on a single multiplexing row due to a failure in the readout. Out of the remaining 407 bolometers, we consider 319 that detect Venus at least 5 times (out of 74 observations) with signal-to-noise (S/N) $>$ 3 and derived end-to-end optical efficiency greater than 5\% (see Section~\ref{sec:calibration}) for further analysis in this paper. In addition to having low optical efficiency, the lower yield of operable W-band detectors in the field can be mostly attributed to three coupled effects: (1) all 74 detectors within a module share a single bias line (compared to maximum 10 detectors per line for Q band and 44 for G band); (2) variations in detector properties within a wafer resulted in variations in their optimal bias points; and (3) a narrower stable bias range prevented accommodating variations in bias point (Effect 2) with a single bias line. We find that the G-band detectors (described in Section \ref{sec:det_G}) do not suffer from the stability issue and can be biased down to $\lesssim$ 10\% $R_\mathrm{N}$. Therefore, for the second W-band instrument, we have implemented various changes in the detector design to closely follow the G-band detector architecture \citep[see][Fig.~2]{dahal2020}. The new W-band detectors are being fabricated at the time of writing.

\subsection{G-band}\label{sec:det_G}
The CLASS G-band instrument, which started observation in September 2019, has a total of 255 dichroic dual-polarization pixels spread among three identical modules \citep{dahal2020}. Each pixel contains four bolometers to simultaneously measure the two linear polarization states at 150 and 220 GHz frequency bands defined through on-chip filtering. These detectors are read out with 24 columns multiplexing 44 rows of SQUIDs. To optimize the detector biasing, each column is provided with a separate bias line. Before the deployment of the G-band instrument, we had reported array yields of 80\% and 57\% for the 150 and the 220~GHz frequency bands, respectively \citep{dahal2020}. For 150~GHz, 389 out of 408 bolometers reported working in the lab detect Jupiter, and for 220 GHz, 211 out of 290 detect Jupiter. The larger difference for the 220 GHz detector array is mostly due to failure of two readout columns during deployment. For this paper, we conservatively exclude those G-band bolometers that detect Jupiter fewer than 5 times (out of 70 observations) with \mbox{S/N $>$ 3}.

\section{Optical Passband and Loading}\label{sec:opt_band_load}
A combination of absorptive, reflective, and scattering filters inside the receiver cryostat \citep{iuliano2018} suppresses the infrared power reaching the focal plane. Due to the details of the detector package geometry \citep{crowe2013}, frequencies below the waveguide cutoff for each design and above the niobium gap energy ($\sim700$~GHz) are suppressed. Between these frequencies, a series of broadband microstrip filters define the band edges and limit the out-of-band spectral response \citep{kpop_2008a, kpop_2008b} for all CLASS detectors. The following subsections describe the measurements and show results for the passbands measured in the lab, and the in-band optical power measured in the field.

\subsection{Frequency Bands}\label{sec:bandpass}
\begin{figure*}
\begin{center}
\includegraphics[scale=0.35]{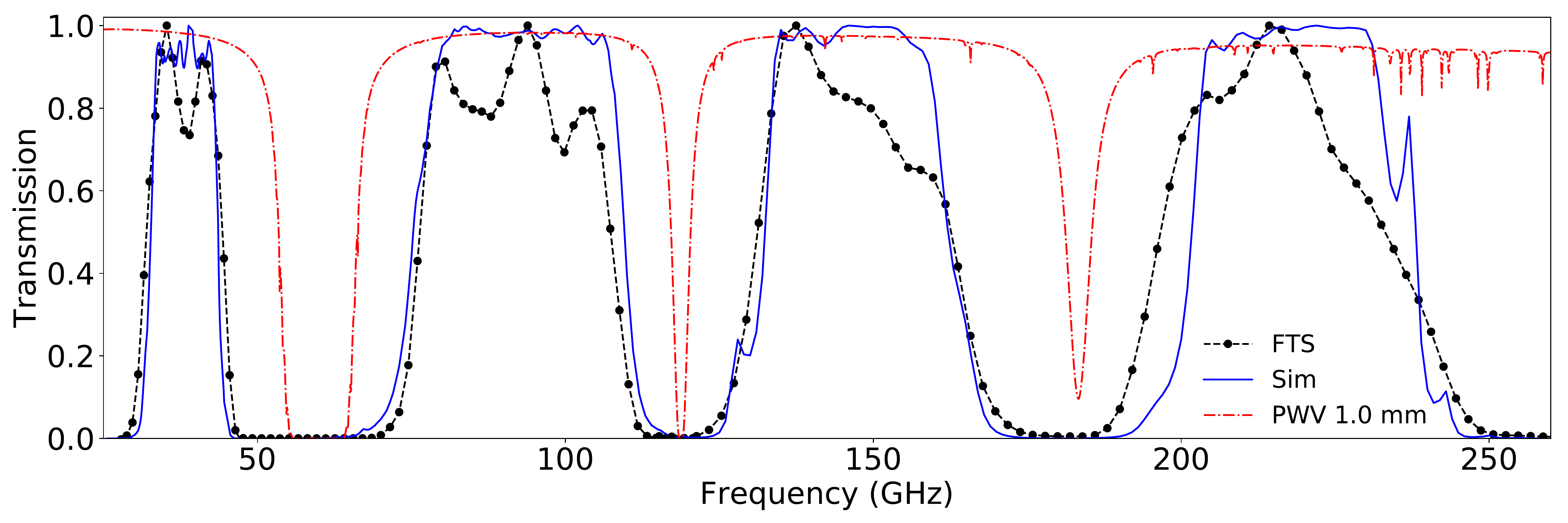}
\end{center}
\caption{Average measured (dotted-black) and simulated (solid-blue) spectral response for different CLASS frequency bands overplotted with the atmospheric transmission model at the CLASS site with PWV = 1 mm (red dashed-dotted). The atmospheric transmission model was obtained from the ALMA atmospheric transmission calculator based on the ATM code described in \cite{padro2001}. The spectral response is normalized to unit peak, and the atmospheric transmission model is in units of fractional power transmitted. The bandwidths and center frequencies for different diffuse sources for these passbands are shown in Table~\ref{tab:bandpass}.}
\label{fig:bandpass}
\end{figure*}

\begin{deluxetable*}{lrrrr}
\tablecaption{\label{tab:bandpass} Measured (and Simulated) Bandwidths and Effective Center Frequencies over CLASS Passbands (in GHz)}
\tablehead{ & \colhead{\textbf{Q-band}} &
\colhead{\textbf{W-band}} & \colhead{\textbf{G-band (Lower)}} & \colhead{\textbf{G-band (Upper)}}
}
\startdata
\textbf{Bandwidths:} \\
FWHP & 12.3 (10.9) $\pm$ 0.9 & 31.0 (34.3) $\pm$ 1.5 & 31.4 (29.7) $\pm$ 2.0 & 36.5 (36.4) $\pm$ 2.0 \\
Dicke & 14.0 (12.1) $\pm$ 0.9 & 34.4 (37.5) $\pm$  1.5 & 37.6 (35.1) $\pm$  2.0 & 47.0 (40.1) $\pm$  2.0 \\
\hline
\textbf{Effective Center Frequencies\tablenotemark{a}:}\\
Synchrotron & 36.8 (37.2) $\pm$  0.5 & 88.7 (88.9) $\pm$  0.8 & 144.2 (145.2) $\pm$  1.0 & 213.6 (217.4) $\pm$  1.0 \\
Rayleigh-Jeans & 38.1 (38.1) $\pm$  0.5 & 91.7 (92.5) $\pm$  0.8 & 146.4 (147.2) $\pm$  1.0 & 216.0 (219.1) $\pm$  1.0 \\
Dust & 38.7 (38.5) $\pm$  0.5 & 93.2 (94.2) $\pm$  0.8 & 147.6 (148.1) $\pm$  1.0 & 217.1 (220.0) $\pm$  1.0 \\
CMB & 38.0 (38.1) $\pm$  0.5 & 91.3 (92.0) $\pm$  0.8 & 145.7 (146.6) $\pm$  1.0 & 214.5 (218.0) $\pm$  1.0 \\
\enddata
\tablenotetext{a}{The values are calculated over CLASS passbands for diffuse sources using Equation \ref{eq:cf}.}
\end{deluxetable*}

We measured the CLASS detector passbands using Martin-Puplett Fourier transform spectrometers (FTSs; \citealt{martin_puplett1970}) in the lab. For the 40 and the 90 GHz detectors, a tabletop FTS made at the Johns Hopkins University with $\sim 1$ GHz resolution \citep{wei_thesis} was used to obtain the spectrum shown in Figure \ref{fig:bandpass}. The passband measurements and FTS testing setup for the 40 and the 90 GHz detectors are described in \citet{appel19} and \citet{dahal2018}, respectively. For the 150 and 220~GHz detectors, the lab cryostat and FTS testing setup did not allow the use of a tabletop FTS; therefore, a smaller and compact FTS with $\sim 2$~GHz resolution \citep{pan2019} was used instead. Figure \ref{fig:bandpass} shows measured and simulated CLASS passbands normalized to unit peak. The measured passbands have been corrected for the transmission through the lab cryostat filters and the frequency-dependent gain for the detector feedhorns that were placed a meter behind a 10-cm diameter cold stop. We co-added measured passbands from a sub-set of detectors with high S/N in each array to obtain the plot shown in Figure \ref{fig:bandpass}. The discrepancies observed in-band between measured and simulated passbands could be due to optical effects in the test setup not included in the simulation like FTS output, coupling between FTS and cryostat, and reflections inside the cryostat. However, the measurements show that all four passbands satisfy the design requirement to safely avoid strong atmospheric emission lines as shown in Figure \ref{fig:bandpass}. 

It is worth noting that, compared to Figure \ref{fig:bandpass}, the passbands for the 150 and the 220~GHz detectors reported in \citet{dahal2020} were shifted higher by a few GHz due to systematics associated with the compact FTS. The G-band measurements presented in \citet{dahal2020} were performed with an FTS setup such that the plane with the wire-grid polarizers in the compact FTS \citep[see][Fig.~1]{pan2019} was parallel to the vertical plane of the CLASS cryostat \citep{iuliano2018}. As the FTS was rotated by 90$^\circ$ aligning the plane with the polarizers to the horizontal plane of the cryostat, the measured passbands shifted lower by a few GHz to produce the result shown in Figure \ref{fig:bandpass}. After the deployment of the G-band instrument, we investigated the FTS systematics in the lab to verify the accuracy of the measured passbands. We performed FTS measurements on spare 90~GHz CLASS detectors (nearly identical to the ones in the field) using a  single-frequency HMC-C030\footnote{\url{www.analog.com/products/hmc-c030}} voltage-controlled oscillator (VCO) source. We tuned the VCO source to 7.5~GHz and used a $\times 12$ frequency multiplier to produce 90 GHz input to the FTS optically coupled to the 90~GHz CLASS detectors inside the cryostat. While the vertical FTS configuration did not produce enough S/N output for the analysis, the horizontal configuration performed as expected showing a peak at $89.8 \pm 1.0$~GHz in response to the 90 GHz input. Therefore, we use FTS measurements from the horizontal configuration  (Figure~\ref{fig:bandpass}) for further G-band analysis in this paper.

For all the measured and simulated passbands, we calculate detector bandwidths in two different ways -- full width at half power (FWHP) and Dicke bandwidth \citep{dicke1946}, which is defined as:
\begin{equation}
    \Delta \nu_{\scriptscriptstyle \mathrm{Dicke}} \equiv \frac{\left[ \int f(\nu) \mathrm{d}\nu \right]^2}{\int f(\nu)^2 \mathrm{d}\nu},
\label{eq:dicke_bw}    
\end{equation}
where $\nu$ is the frequency and $f(\nu)$ is the spectral response. Following \cite{page2003b}, we also calculate the effective central frequencies for the measured passbands as: 

\begin{equation}
    \nu_\mathrm{e} \equiv \frac{\int\nu f(\nu) \sigma(\nu) \mathrm{d}\nu}{\int f(\nu) \sigma(\nu) \mathrm{d}\nu},
\label{eq:cf}    
\end{equation}

where $\sigma (\nu)$ describes the frequency dependence for different sources. For a beam-filling Rayleigh-Jeans (RJ) source, the detector has a flat spectral response as the source spectrum is exactly canceled by the single-moded throughput; therefore, we set \mbox{$\sigma (\nu)=1$}. For the diffuse synchrotron and dust sources, we use $\sigma (\nu)\propto\nu^{-3.1}$ and $\nu^{1.55}$, respectively \citep{planck2016-l04}. For the CMB, since the source measured is the anisotropy, we set:

\begin{equation}
   \sigma (\nu) \propto \left. \frac{1}{\nu^2} \frac{\partial B(\nu, T)}{\partial T} \right|_{T=T_\mathrm{cmb}} \propto 
   \frac{\nu^2\mathrm{e}^x}{(\mathrm{e}^x-1)^2}\ ,
\label{eq:sigmanu_cmb}
\end{equation}

where $B(\nu, T)$ is the Planck blackbody, $T_\mathrm{cmb} = 2.725$~K, and $x = h\nu/kT_\mathrm{cmb}$, where $h$ and $k$ are the Planck and the Boltzmann constants, respectively. The calculated bandwidths and effective central frequencies for all these diffuse sources for both the measured and the simulated passbands are tabulated in Table~\ref{tab:bandpass}. The associated uncertainties
are the quadrature summations of the respective standard errors on the mean and their measurement resolutions (full FTS resolution for bandwidths and half the resolution for effective center frequencies). The latter dominates the quoted uncertainties for all four frequency bands.

\subsection{Optical Loading}\label{sec:loading}
\begin{figure}[t]
    \begin{center}
    \includegraphics[scale=0.3]{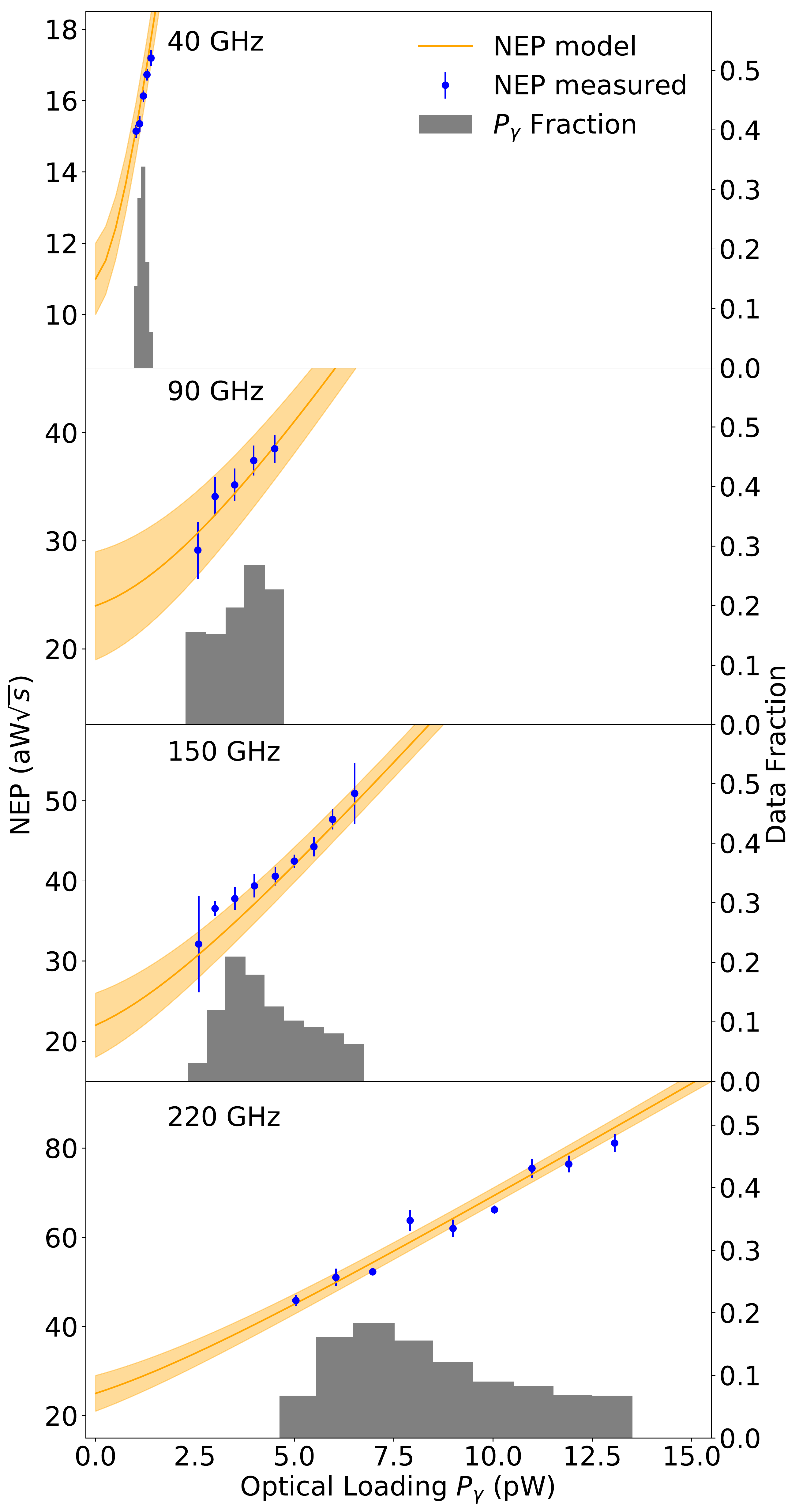}
    \caption{Array-averaged NEP vs $P_\gamma$ for different CLASS frequency bands. The blue points are the measured values from CMB survey scans whereas the orange curves are NEP models (Equation \ref{eq:nep}) based on lab-measured detector parameters. The error bars on the data points and the shaded regions around the models are the 1$\sigma$ uncertainties. Before binning the measured NEP values, a data selection cut of wind speed $<$ 1.5 $\mathrm{m\, s^{-1}}$ was applied for all four detector arrays. The histograms show the spread of array-averaged $P_\gamma$ measured during the entire Era 2 observing campaign. The measured NEP values from CMB scans are consistent with the models at all four frequency bands.}
    \label{fig:nep_load}
     \end{center}
\end{figure}

The CLASS observation strategy is to scan azimuthally across 720$^\circ$ at a constant elevation of 45$^\circ$ \citep{xu2020}. The telescope boresight angle is changed by $15^\circ$ once per 24-hour observing cycle, nominally covering seven boresight angles from $-45^\circ$ to $+45^\circ$ each week.
At the beginning of the observing cycle for the day, we acquire $I$-$V$ curves to select the optimal detector bias voltage. For W-band, we apply one voltage bias per module (i.e., four columns), whereas for Q- and G-band, we choose one bias per column. Using the $I$-$V$ data, we measure the detector bias power ($P_\mathrm{bias}$) defined as $P_\mathrm{bias} = I \times V$ at 80\% $R_\mathrm{N}$. The detector optical loading $P_\gamma$ can then be calculated by subtracting this $P_\mathrm{bias}$ from the detector saturation power $P_\mathrm{sat}$ calculated during dark lab tests, i.e., $P_\gamma = P_\mathrm{sat} - P_\mathrm{bias}$. We show the spread of array-averaged $P_\gamma$ for all four frequency bands during the \mbox{Era 2} observing campaign in Figure~\ref{fig:nep_load}. The range of measured optical loading narrows substantially at the lower frequency bands, highlighting the stability of the atmospheric emission at the CLASS site. The prominent atmospheric features near the CLASS frequency bands are the 60 and 117~GHz oxygen emission lines and the 183~GHz water emission line (Figure~\ref{fig:bandpass}). Therefore, due to their proximity to the water line, the loading at the higher frequency bands increases with the precipitable water vapor (PWV) levels. Since the Q band is the furthest away from the 183~GHz water line, its loading is least sensitive to weather variations as the oxygen emission remains relatively constant. The median array-averaged loading $P_\gamma$ during the observing campaign for the 40, 90, 150, and 220 GHz detector arrays were 1.2, 3.6, 4.4, and 8.5 pW, respectively. Note that while Figure \ref{fig:nep_load} shows the data for the 40~GHz instrument with the TG filter installed, the median loading without the filter was the same.

\section{Noise Performance}\label{sec:noise}
CLASS detectors are background limited, meaning that the optical loading drives the total detector noise measured. For a given $P_\gamma$, the noise-equivalent power (NEP) for a detector can be modeled as:

\begin{equation}
\mathrm{NEP}^2 = \mathrm{NEP}^2_\mathrm{d} + h\nu_0 P_\gamma + \frac{P^2_\gamma}{\Delta\nu},
\label{eq:nep}    
\end{equation}

where NEP$_\mathrm{d}$ is the dark detector noise, $h$ is the Planck constant, and $\nu_0$ and $\Delta\nu$ are the detector center frequency and bandwidth, respectively. In Figure \ref{fig:nep_load}, we compare the measured NEP from on-sky data to the NEP model from Equation \ref{eq:nep} for all four CLASS frequency bands. For the NEP model, we use lab-measured RJ center frequencies and FWHP bandwidths from Table~\ref{tab:bandpass}, and NEP$_\mathrm{d}$ of \mbox{11 $\pm$ 1 aW$\sqrt{\mathrm{s}}$} \citep{appel14}, \mbox{24 $\pm$ 5 aW$\sqrt{\mathrm{s}}$} \citep{dahal2018}, \mbox{22 $\pm$ 4 aW$\sqrt{\mathrm{s}}$}, and \mbox{25 $\pm$ 4 aW$\sqrt{\mathrm{s}}$} \citep{dahal2020} for the 40, 90, 150, and 220 GHz arrays, respectively. The model uncertainty is based on the errors measured in these parameters. The measured NEP values are computed from the PSD of pair-differenced 10-minute TODs obtained from on-sky observations. First, we subtract detector TODs from each polarization pair within a pixel to reduce any correlated noise. Then, we take the average of the PSD in the side-bands of the 10~Hz modulation frequency, and divide the average by two to recover the per-detector NEP. For all four detector arrays, we average the PSDs over the side-bands of 8.0~--~9.0~Hz and 11.0~--~12.0~Hz.

Figure \ref{fig:nep_load} shows the binned pair-differenced NEP averaged across the array vs the average $P_\gamma$ for all four CLASS frequency bands. Since the $P_\gamma$ values are based on the $I$-$V$ measurements and we only acquire $I$-$V$ data once per day during nominal CMB scans, we only bin the NEP values from the TODs acquired within four hours after an $I$-$V$ is acquired. This is especially important for the G-band detectors as they can have larger variations in the atmospheric loading throughout the day. It is also worth noting that although single detectors without an operational pair were not included in this analysis, they could still be mapped for the CMB analysis.

Before binning the NEP vs $P_\gamma$ values shown in Figure~\ref{fig:nep_load}, we applied a data selection cut for all four detector arrays to remove the TODs acquired when the wind speed at the site was higher than 1.5 $\mathrm{m\, s^{-1}}$. The effect of high wind speeds on CLASS data is discussed in \cite{harrington2021} and will be explored further in upcoming publications. In Section \ref{sec:cmb_sens}, we compute the overall sensitivity for all CLASS detector arrays without applying any wind speed cuts to the TODs. For Figure~\ref{fig:nep_load}, a wind speed $<$~1.5~$\mathrm{m\, s^{-1}}$ data selection cut was applied to compare the data to the detector NEP model.

For all four frequency bands, the on-sky measured NEP values are consistent with the NEP models based on lab-measured parameters as shown in Figure \ref{fig:nep_load}. The measured NEPs are dominated by photon bunching (third term in Equation \ref{eq:nep}) and verify that the CLASS detectors are background-limited at all four frequency bands, as designed. Compared to Era 1 observations \citep[see][Fig.~3]{appel19}, the NEP for 40~GHz detectors decreased by $\sim$ 21\%, primarily due to lower optical loading achieved through the Q-band instrument upgrade (Section \ref{sec:det_Q}).

\section{Moon and Planet Observations}\label{sec:planet}
CLASS periodically performs dedicated observations of the Moon, Venus, and Jupiter to calibrate the detectors' power response, obtain telescope pointing information, and characterize the beam response. During these dedicated observations, the telescopes scan across the source azimuthally at a fixed elevation as the source rises or sets through the telescopes' fields of view. Since the Moon provides the highest S/N, it is used to obtain pointing information \citep{xu2020} for all CLASS telescopes. However, it saturates the W-band and G-band detectors, and its absolute Q-band brightness temperature is not well established. Therefore, when available, we use Venus and/or Jupiter observations to obtain detector calibration from measured power to antenna temperature of the source.

During Era 2, all three CLASS telescopes performed 70 dedicated Jupiter observations. In addition, the Q- and W-band telescopes performed 74 dedicated Venus scans. Since the G-band receiver was deployed, Venus has not been available for observations. Therefore, for the detector calibration summarized in Section \ref{sec:calibration}, we use Jupiter observations for 150 and 220 GHz detectors whereas we use average values obtained from both Venus and Jupiter for 40 and 90 GHz detectors. After the TG filter was installed in the Q-band instrument, neither planet has been available for observations; hence we use Moon observations instead for this particular configuration.

\subsection{Calibration to Antenna Temperature}\label{sec:calibration}
Given the telescope beam sizes (see Table \ref{tab:opt_summary}), both planets (and the Moon for Q band) are well approximated as ``point sources'' whose brightness temperature ($T_\mathrm{p}$) relates to the peak response measured by CLASS detectors ($T_\mathrm{m}$) as:

\begin{equation}
    T_\mathrm{p} \Omega_\mathrm{p} = T_\mathrm{m} \Omega_\mathrm{B},
\label{eq:scaling}
\end{equation}

where $\Omega_\mathrm{B}$ is the beam solid angle and $\Omega_\mathrm{p}$ is the solid angle subtended by the source \citep{page2003a}. For W- and G-band detectors, $T_\mathrm{m}$ is also corrected for atmospheric transmission to account for the effect of PWV at the CLASS site. We calculate the correction factor using the PWV data from APEX\footnote{\url{https://archive.eso.org/wdb/wdb/asm/meteo\_apex/form}} and ACT \citep{bustos2014} radiometers along with the atmospheric transmission model based on \citet{padro2001} shown in Figure~\ref{fig:bandpass}. No correction was necessary for the Q-band detectors as the effect of PWV on $T_\mathrm{m}$ was less than a percent.

For a given CLASS detector, the calibration factor from power deposited on the bolometer $\mathrm{d}P_\gamma$ to antenna (RJ) temperature on the sky $\mathrm{d}T_\mathrm{RJ}$ can now be written as:

\begin{equation}
    \frac{\mathrm{d}T_\mathrm{RJ}}{\mathrm{d}P_\gamma} = \frac{T_\mathrm{m}}{P_0}= \frac{T_\mathrm{p}}{P_0} \frac{\Omega_\mathrm{p}}{\Omega_{\mathrm{B}}},
\label{eq:calib_rj}
\end{equation}

where $P_0$ is the peak power amplitude measured by the bolometer. For Jupiter observations, we use $T_\mathrm{p}$ =  152.6~$\pm$~0.6~K,\footnote{Considering the steep Jupiter spectrum at Q band, we use a local power law fit between WMAP's Ka and Q-band measurements to correct for $\sim$~2~GHz difference in effective center frequencies between WMAP and CLASS Q-band detectors.} 172.8~$\pm$~0.5~K \citep{bennett2013}, 174.1~$\pm$~0.9~K, and 175.8~$\pm$~1.1~K \citep{planck2016-LII} for the 40, 90, 150, and 220 GHz detectors, respectively. For Venus observations, we use $T_\mathrm{p}$ = 430.4~$\pm$~2.8~K for 40~GHz and 354.5~$\pm$~1.3~K for 90~GHz from \citet{dahal2020_venus}.\footnote{We note that these Venus brightness temperature values were obtained from a separate analysis using CLASS data and WMAP-measured Jupiter brightness temperatures. Refer to \citet{dahal2020_venus} for details.} For the 40 GHz Moon observations used to calibrate the TG-filter configuration, we use \mbox{$T_\mathrm{p} = 200 \pm 1$~K}. Following a similar procedure to that described in \citet{dahal2020_venus}, this value was obtained by comparing Moon to Jupiter observations performed when the TG filter was not in place.  

Finally, we obtain per-detector $\mathrm{d}T_\mathrm{RJ}/\mathrm{d}P_\gamma$ values by averaging all the individual observations of a source acquired throughout the observing campaign. This per-detector averaging is performed on a map level \citep{xu2020}. For every dedicated planet/Moon observation, a source-centered map is produced using the TOD and telescope pointing information for a given detector. The individual maps are then averaged to form an aggregate map of the source from which $P_0$ and $\Omega_\mathrm{B}$ values for Equation \ref{eq:calib_rj} are extracted. Since $\Omega_\mathrm{p}$ changes between observations, the detector response is scaled by 1/$\Omega_\mathrm{p}$ while averaging the maps.
While $\Omega_\mathrm{p}$ of all three sources vary with the changing distance between the source and CLASS telescopes, $\Omega_\mathrm{p}$ of Jupiter also varies with time due to changes in viewing angle for the oblate planet. We correct for this Jupiter disk oblateness using the method described in \citet{weiland11}. In addition, for the Q-band Moon observations, the detector response is scaled by $\langle T_\mathrm{model} \rangle / T_\mathrm{model}$, where $T_\mathrm{model}$ is a phase-dependent disk-averaged Moon brightness temperature model \citep{xu2020} derived from the Chang'E 37~GHz data \citep{zheng2012} and $\langle T_\mathrm{model} \rangle $ is the model brightness temperature averaged over all Moon phases.

\begin{figure}[ht]
\begin{center}
\includegraphics[scale=0.37]{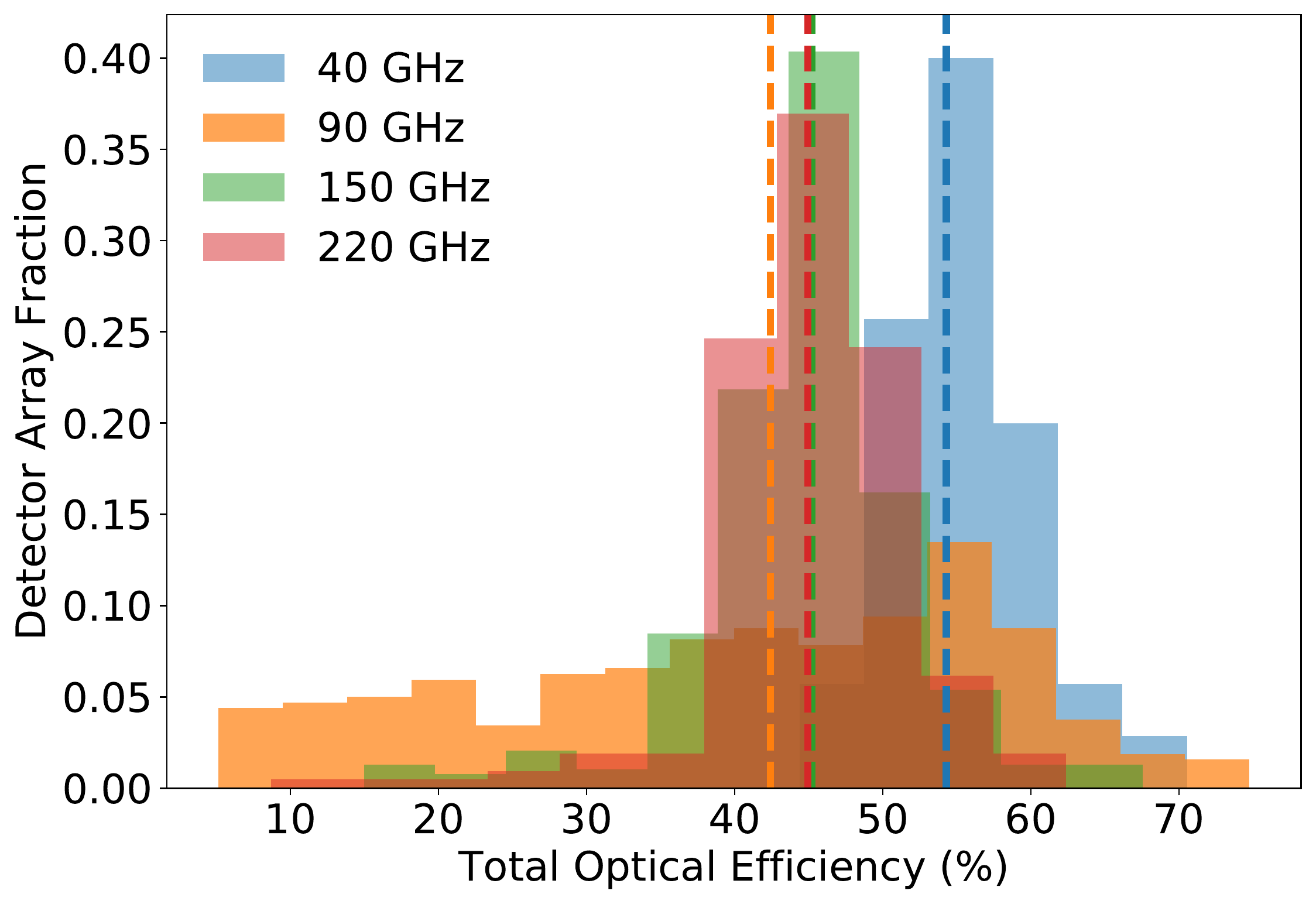}
\end{center}
\caption{The distribution of end-to-end (detector plus telescope optics) optical efficiency  for CLASS detector arrays obtained from planet observations. The
dashed lines represent the array medians, and the array fraction is based on the total number of optically-sensitive detectors (N$_\mathrm{det}$ in Table \ref{tab:opt_summary}) in a given array. The 40 GHz efficiency values shown here were obtained after the April 2018 receiver upgrade and without the TG filter installed; the TG filter lowers the median efficiency to 43\%.}
\label{fig:eff_dist}
\end{figure}

\begin{deluxetable*}{lcrrrr}[ht]
\tablecaption{\label{tab:opt_summary} Optical Performance Summary (Array Medians) of CLASS Telescopes}
\tablehead{
\colhead{\textbf{Parameter}} & \colhead{\textbf{Symbol [Unit]}} & \colhead{\textbf{40 GHz\tablenotemark{a}}} & \colhead{\textbf{90 GHz}} & \colhead{\textbf{150 GHz}} & \colhead{\textbf{220 GHz}}
}
\startdata
Beam FWHM & $\theta$ [arcmin] & 95 & 37 & 23 & 16 \\   
Beam Solid Angle & $\Omega$ [$\mathrm{\upmu}$sr] & 822 & 136 & 51 & 26 \\
Telescope Efficiency & $\eta$ & 0.54 (0.43) & 0.42 & 0.45 & 0.45 \\
RJ Temperature Calibration & $\frac{\mathrm{d}T_\mathrm{RJ}}{\mathrm{d}P_\gamma}$ [$\mathrm{K \, pW^{-1}}$] & 10.8 (13.8)  & 5.5 & 5.1 & 4.4\\
CMB-RJ Calibration & $\frac{\mathrm{d}T_\mathrm{cmb}}{\mathrm{d}T_\mathrm{RJ}}$ & 1.04 & 1.23 & 1.68 & 2.92 \\
Detector Dark Noise Power & NEP$_\mathrm{d}$ [aW$\sqrt{\mathrm{s}}$] & 11 & 24 & 22 & 25 \\
Detector Total Noise Power & NEP [aW$\sqrt{\mathrm{s}}$]  & 16 (15) & 47 & 51 & 83 \\
Detector Noise Temperature & NET [$\mathrm{\upmu}$K$_\mathrm{cmb}\sqrt{\mathrm{s}}$] & 180 (217)  & 346  & 453 &  1034 \\
No. of Optical Detectors & N$_\mathrm{det}$ & 70 & 319 & 389 & 211 \\
Array Noise Temperature & NET$_\mathrm{array}$ [$\mathrm{\upmu}$K$_\mathrm{cmb}\sqrt{\mathrm{s}}$] & 22 (26)  & 19 & 23 & 71 \\
\enddata
\tablenotetext{a}{40 GHz optical performance after April 2018 upgrade. The values in parentheses correspond to the telescope's performance with the TG filter installed.}
\end{deluxetable*}

Table \ref{tab:opt_summary} shows the array median $\mathrm{d}T_\mathrm{RJ}/\mathrm{d}P_\gamma$ calibration factors for all four CLASS frequency bands. These calibration factors relate to the telescopes' optical efficiency ($\eta$) as follows:

\begin{equation}
\label{eq:eff}
\begin{split}
    \eta = \left(k\Delta\nu\frac{\mathrm{d}T_\mathrm{RJ}}{\mathrm{d}P_\gamma}\right)^{-1} 
    &= 0.54^{+0.05}_{-0.03} \ (40 \ \mathrm{GHz} \mathrm{\  w/o\ TG}) \\
    &= 0.43^{+0.03}_{-0.03} \ (40 \ \mathrm{GHz} \mathrm{\  w/\ TG}) \\
    &= 0.42^{+0.15}_{-0.22} \ (90 \ \mathrm{GHz}) \\
    &= 0.45^{+0.05}_{-0.06} \ (150\  \mathrm{GHz}) \\
    &= 0.45^{+0.05}_{-0.04} \ (220\ \mathrm{GHz}),
\end{split}
\end{equation}

where $k$ is the Boltzmann constant and $\Delta\nu$ is the detector bandwidth (measured FWHP) from Table \ref{tab:bandpass}. The values in Equation \ref{eq:eff} are array medians, and the uncertainties indicate 68\% widths of the respective distributions. Figure \ref{fig:eff_dist} shows the distribution of $\eta$ for all four CLASS frequency bands. Compared to Era 1 of CLASS observations \citep{appel19}, we achieved $\sim$ 7\% higher optical efficiency at 40~GHz (without TG filter) by removing the MMFs from the Q-band receiver (Section \ref{sec:det_Q}). The median optical efficiencies achieved by 90, 150, and 220~GHz detector arrays are similar to that of the 40~GHz array in Era~1. For the 90~GHz array, while some detectors have efficiencies similar to that of the other frequency bands, the spread is large and skewed toward low efficiency as shown in Figure \ref{fig:eff_dist}. We are actively investigating the cause of this larger efficiency spread by revisiting the design of the W-band detector components.

\subsection{CMB Sensitivity}\label{sec:cmb_sens}
The antenna temperature $\mathrm{d}T_\mathrm{RJ}$ can be converted to the CMB thermodynamic temperature $\mathrm{d}T_\mathrm{cmb}$ as:

\begin{equation}
    \frac{\mathrm{d}T_\mathrm{cmb}}{\mathrm{d}T_\mathrm{RJ}} \approx \frac{(\mathrm{e}^{x_0}-1)^2}{x_0^2\mathrm{e}^{x_0}},
\label{eq:calib_cmb}
\end{equation}

where $x_0 = h\nu_0/kT_\mathrm{cmb}$ and $\nu_0$ is the CMB center frequency from Table \ref{tab:bandpass}. Combining Equations \ref{eq:calib_rj} and \ref{eq:calib_cmb}, we can calculate the calibration factor from $\mathrm{d}P_\gamma$ to $\mathrm{d}T_\mathrm{cmb}$ as follows:

\begin{equation}
\begin{split}
    \frac{\mathrm{d}T_\mathrm{cmb}}{\mathrm{d}P_\gamma} &= \frac{\mathrm{d}T_\mathrm{cmb}}{\mathrm{d}T_\mathrm{RJ}} \frac{\mathrm{d}T_\mathrm{RJ}}{\mathrm{d}P_\gamma} \\
    &= 11.3^{+0.7}_{-0.9} \  \mathrm{K \, pW}^{-1}  \ (40 \ \mathrm{GHz} \mathrm{\  w/o\ TG}) \\
    &= 14.4^{+1.3}_{-1.0}  \  \mathrm{K \, pW}^{-1}  \ (40 \ \mathrm{GHz} \mathrm{\  w/\ TG}) \\
    &= 6.8^{+7.7}_{-1.8} \  \mathrm{K \, pW}^{-1} \ (90 \ \mathrm{GHz}) \\
    &= 8.6^{+1.3}_{-0.9} \  \mathrm{K \, pW}^{-1} \ (150\  \mathrm{GHz}) \\
    &= 12.9^{+1.3}_{-1.3} \  \mathrm{K \, pW}^{-1} \ (220\ \mathrm{GHz}),
\end{split}
\label{eq:calib_final}
\end{equation}

where the values are array medians, and the uncertainties are 68\% widths of the respective distributions. While we intend to calibrate the CLASS CMB maps through cross-correlation with maps from WMAP and \textit{Planck}, the $\mathrm{d}T_\mathrm{cmb}/\mathrm{d}P_\gamma$ in Equation \ref{eq:calib_final} obtained from planets/Moon can be used to check CLASS noise modeling to achieve additional constraints on experiment characterization.

Multiplying the per-detector $\mathrm{d}T_\mathrm{cmb}/\mathrm{d}P_\gamma$ by the NEP values (Section \ref{sec:noise}) gives the detector's CMB sensitivity in terms of noise-equivalent temperature (NET). For every detector, we first calculate a median NEP from all the 10-minute TODs acquired throughout the observing campaign. Using the per-detector median NEP and $\mathrm{d}T_\mathrm{cmb}/\mathrm{d}P_\gamma$, we obtain the detector NET values, which are summarized in Table \ref{tab:opt_summary}. Finally, we calculate the array noise temperatures by inverse-variance weighting the NETs of all the optically-sensitive detectors in a given array. These array noise temperatures presented in Table~\ref{tab:opt_summary} represent instantaneous array sensitivities achieved by CLASS telescopes during the Era 2 observing campaign. In terms of sky maps, these noise temperatures refer to white noise amplitudes, which contribute to maps at all angular scales. Future publications will assess additional noise sources impacting larger angular scales.

\section{Conclusion}\label{sec:conclusion}
We have presented the on-sky performance of the CLASS receivers at 40, 90, 150, and 220 GHz frequency bands based on data acquired during the Era 2 observing campaign between April 2018 and March 2020. In this paper, we summarize the measured detector parameters including optical loading, time constants, and passbands, and verify that the on-sky NEP measurements are consistent with expected detector noise models. Using Moon, Venus, and Jupiter as calibrators, we obtain end-to-end optical efficiency and measured power to antenna temperature calibration for all the optically-sensitive detectors. These measurements imply instantaneous array sensitivities of 22, 19, 23, and 71 $\mathrm{\upmu}$K$_\mathrm{cmb}\sqrt{\mathrm{s}}$ for the 40, 90, 150, and 220 GHz detector arrays, respectively. As a result of the instrument upgrade, the 40 GHz telescope is now $\sim$ 31\% more sensitive than it was during Era 1 observations \citep{appel19}. The addition of a second 90 GHz instrument will further increase CLASS's CMB sensitivity near the minimum of polarized Galactic emission.

This paper is a part of a series of publications on the CLASS Era 2 multi-frequency observations. Upcoming publications will address beam profile, polarization performance, and science results.

\section*{ACKNOWLEDGMENTS}
We acknowledge the National Science Foundation Division of Astronomical Sciences for their support of CLASS under Grant Numbers 0959349, 1429236, 1636634, 1654494, and 2034400. We thank Johns Hopkins University President R. Daniels and the Deans of the Kreiger School of Arts and Sciences for their steadfast support of CLASS. We further acknowledge the very generous support of Jim and Heather Murren (JHU A\&S ’88), Matthew Polk (JHU A\&S Physics BS ’71), David Nicholson, and Michael Bloomberg (JHU Engineering ’64). The CLASS project employs detector technology developed in collaboration between JHU and Goddard Space Flight Center under several previous and ongoing NASA grants. Detector development work at JHU was funded by NASA cooperative agreement 80NSSC19M0005. CLASS is located in the Parque Astron\'omico Atacama in northern Chile under the auspices of the Agencia Nacional de Investigaci\'on y Desarrollo (ANID).

We acknowledge scientific and engineering contributions from Max Abitbol, Fletcher Boone, Lance Corbett, David Carcamo, Mauricio D\'iaz, Pedro Fluxa, Dominik Gothe, Ted Grunberg, Mark Halpern, Saianeesh Haridas, Connor Henley, Gene Hilton, Johannes Hubmayr, Ben Keller, Lindsay Lowry, Nick Mehrle, Grace Mumby, Diva Parekh, Bastian Pradenas, Isu Ravi, Carl Reintsema, Daniel Swartz, Emily Wagner, Bingjie Wang, Qinan Wang, Tiffany Wei, Zi\'ang Yan, Lingzhen Zeng, and Zhuo Zhang. For essential logistical support, we thank Jill Hanson, William Deysher, Miguel Angel D\'iaz, Mar\'ia Jos\'e Amaral, and Chantal Boisvert. We acknowledge productive collaboration with Dean Carpenter and the JHU Physical Sciences Machine Shop team.

S.D. is supported by an appointment to the NASA Postdoctoral Program at the NASA Goddard Space Flight Center, administered by Universities Space Research Association under contract with NASA. S.D. acknowledges support under NASA-JHU Cooperative Agreement 80NSSC19M005. K.H. is supported by NASA under award number 80GSFC17M0002. R.R. acknowledges partial support from CATA, BASAL grant AFB-170002, and 
CONICYT-FONDECYT through grant 1181620. Z.X. is supported by the Gordon and Betty Moore Foundation. 

\software{\texttt{PyEphem} \citep{rhodes2011}, \texttt{NumPy} \citep{numpy}, \texttt{SciPy} \citep{scipy}, \texttt{Astropy} \citep{astropy}, \texttt{Matplotlib} \citep{matplotlib}}

\bibliography{ref, Planck_bib}{}
\bibliographystyle{aasjournal}
\end{document}